\documentclass[draft,letterpaper]{aipproc}
\layoutstyle{6x9}
\usepackage{epsfig}

\begin{document}

\title[\textbf{cond-mat/0211175}]
  {Statistical Mechanics of Money, Income, and Wealth: 
  A Short Survey}

\author{Adrian A.\ Dr\u{a}gulescu}
  {address={Constellation Energy Group, Baltimore, USA}}
\author{Victor M.\ Yakovenko}
  {address={Department of Physics, University of Maryland, 
     College Park, MD 20742-4111, USA}}

\date{10 November 2002}

\begin{abstract}
In this short paper, we overview and extend the results of our papers
\cite{DY-money,DY-income,DY-wealth}, where we use an analogy with
statistical physics to describe probability distributions of money,
income, and wealth in society.  By making a detailed quantitative
comparison with the available statistical data, we show that these
distributions are described by simple exponential and power-law
functions.
\end{abstract}
\maketitle

The equilibrium statistical mechanics is based on the Boltzmann--Gibbs
law, which states that the probability distribution function (PDF) of
energy $\varepsilon$ is $P(\varepsilon)=Ce^{-\varepsilon/T}$, where
$T$ is the temperature, and $C$ is a normalizing constant.  The main
ingredient in the textbook derivation of the Boltzmann-Gibbs law is
conservation of energy.  Similarly, when two economic agents make a
transaction, some amount of money is transferred from one agent to
another, but the sum of their monies before and after transaction is
the same: $m_1+m_2=m_1'+m_2'$.  So, money is locally conserved in
interactions between agents.  Then, by analogy with statistical
physics, one may expect that the equilibrium PDF of money $m$ in a
closed system of agents has the Boltzmann-Gibbs form
$P(m)=e^{-m/T}/T$, where $T$ is the effective ``money temperature''
equal to the average amount of money per agent.  This conjecture was
confirmed in computer simulations of various simple models of money
exchange in Ref.\ \cite{DY-money} under quite general conditions of
the time-reversal symmetry and sharp boundary conditions at the lower
end of $m$.  In a more general case, where the time-reversal symmetry
is broken or debt is permitted \cite{Redner}, the probability
distribution of money may deviate from the Boltzmann-Gibbs law.  A
popular review of these models can be found in Ref.\ \cite{Hayes}.

\begin{figure}
\centerline{
   \epsfig{file=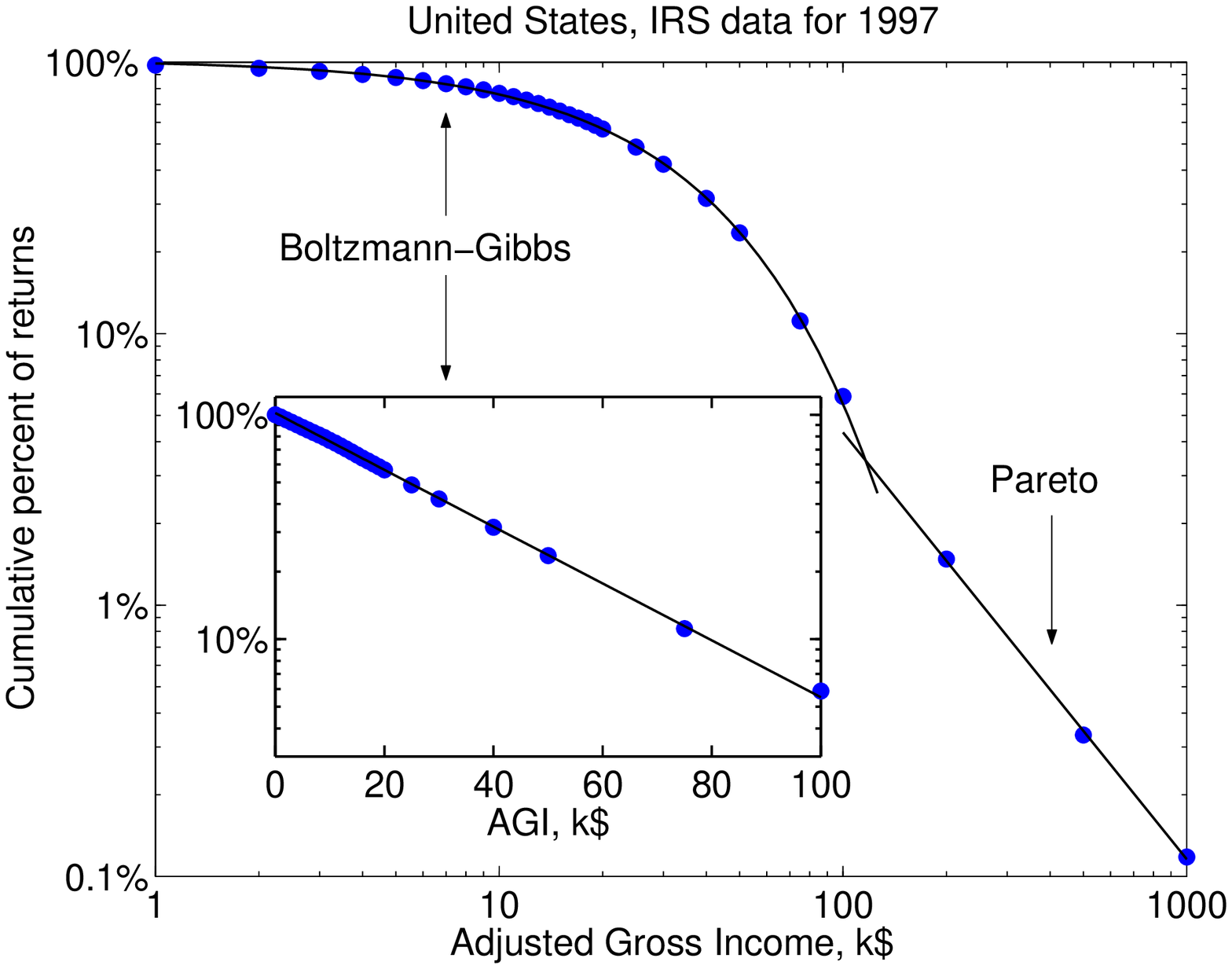,width=0.545\linewidth}
   \epsfig{file=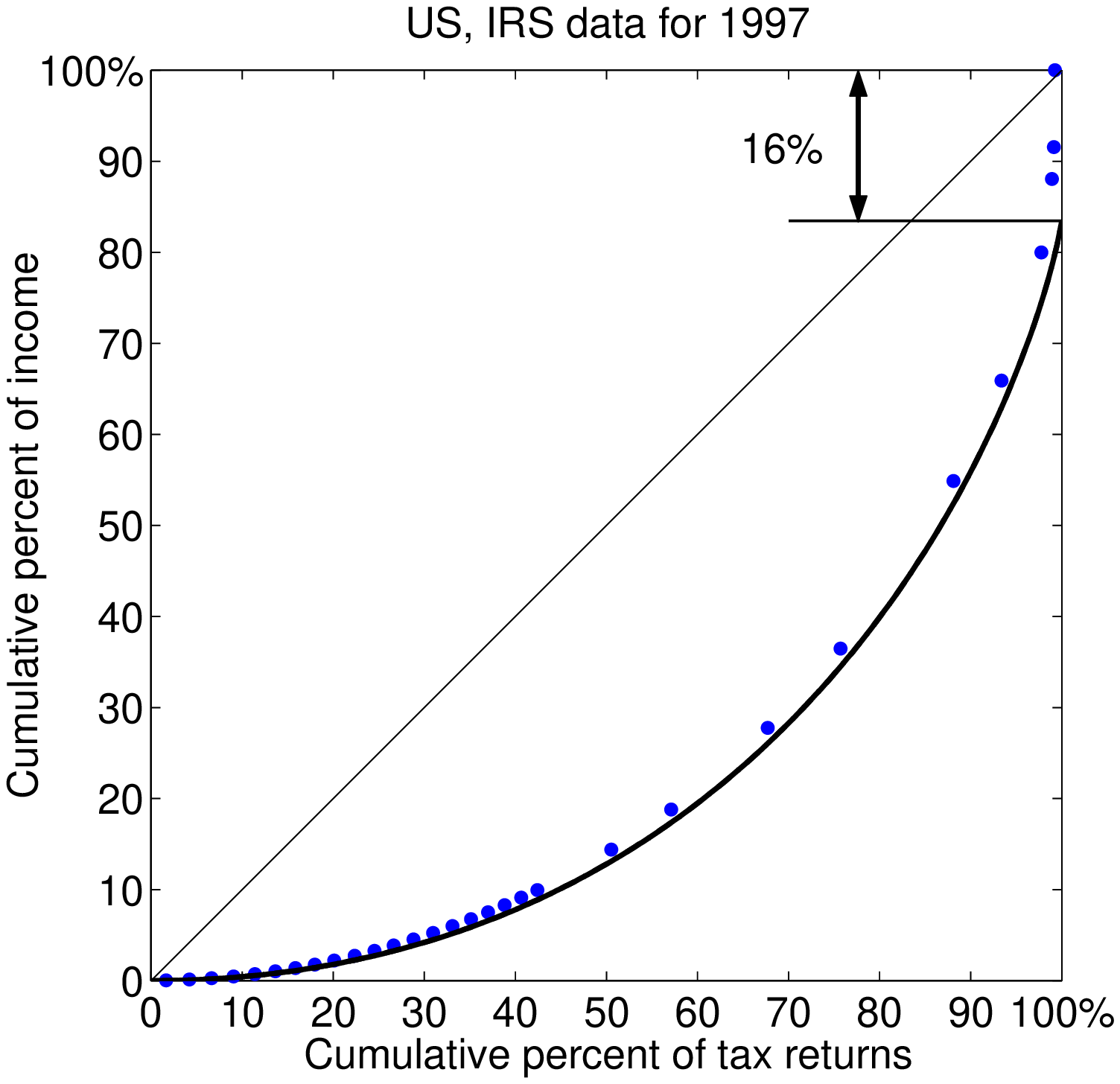,width=0.445\linewidth}}
\caption{ Left panel: Cumulative probability distribution of
  individual income in USA in 1997, shown in log-log scale (main
  panel) and log-linear scale (inset).  The points are the raw data
  from IRS, and the solid lines are the fits to exponential and
  power-law functions.  Right panel: Lorenz plot of the same data
  points, compared with function (\ref{iLorenzBose}) shown by the
  solid line.}
\label{fig:irs97G+P}
\end{figure}

It would be very interesting to compare these results with the actual
PDF of money in the society.  Unfortunately, we were not able to find
such data.  On the other hand, we found a lot of statistical data for
the PDF of income $r$.  In Fig.\ \ref{fig:irs97G+P}, we show the IRS
data for the distribution of individual income in USA in 1997
\cite{Pub1304}.  The left panel shows the cumulative PDF
$N(r)=\int_r^\infty P(r')\, dr'$, which gives the fraction of
individuals with income greater than $r$.  The main panel shows the
data in the log-log scale, and the inset in the log-linear scale.  The
straight line in the inset demonstrates that, for incomes below 100
k\$/year, the income PDF has the exponential Boltzmann-Gibbs form
$P_1(r)=e^{-r/R}/R$, where $R$ is the effective ``income temperature''
equal to the average income.  On the other hand, for very high incomes
above 100 k\$/year, the PDF changes to the Pareto power law, as
illustrated by the straight line in the log-log scale.  The fraction
of population in the power-law tail is very small, less than 3\%.  So,
the income distribution of the great majority of population is
described by the exponential Boltzmann-Gibbs law.

Another standard way of representing income distribution is the
so-called Lorenz plot shown in the right panel of Fig.\
\ref{fig:irs97G+P}.  The horizontal axis of the Lorenz curve, $x(r)$,
represents the cumulative fraction of population with incomes below
$r$, and the vertical axis $y(r)$ represents the fraction of the total
income this population accounts for:
\begin{equation} \label{eq:xy}
  x(r)=\int_0^r P(r')\,dr',\quad
  y(r)=\frac{\int_0^r r' P(r')\,dr'}{\int_0^\infty r' P(r')\,dr'}.
\end{equation}
As $r$ changes from 0 to $\infty$, $x$ and $y$ change from 0 to 1, and
Eq.\ (\ref{eq:xy}) parametrically defines the Lorenz curve in the
$(x,y)$ space.  The diagonal line $y=x$ represents the Lorenz curve in
the case where all population has equal income.  The inequality of the
actual income distribution is measured by the Gini coefficient $0\leq
G\leq1$, which is the area between the diagonal and the Lorenz curve,
normalized to the area of the triangle beneath the diagonal:
$G=2\int_0^1(x-y)\,dx$.

For the exponential PDF $P_1(r)$, the Lorenz curve is
$y=x+(1-x)\ln(1-x)$, and the Gini coefficient is $G_1=1/2$
\cite{DY-income}.  As shown in Fig.\ 3 of Ref.\ \cite{DY-income},
$G_1=1/2$ is in overall good agreement with the Gini coefficient given
by IRS for the last 20 years.  However, because the PDF shown in the
left panel of Fig.\ \ref{fig:irs97G+P} is a mix of exponential and
power-law functions, the Lorenz curve is modified as follows:
\begin{equation} \label{iLorenzBose}
   y=(1-f)[x + (1-x)\ln(1-x)] + f\delta(1-x).
\end{equation}
In Eq.\ (\ref{iLorenzBose}), the weight of the first term is reduced
by $1-f$, because the normalization factor for $y$ in Eq.\
(\ref{eq:xy}) differs from the purely exponential case.  The remaining
weight $f$ is the fraction of the total income contained in the
power-law tail in excess of the exponential law.  Because the fraction
of population in the tail is very small, this contribution is
approximated by the delta-function in Eq.\ (\ref{iLorenzBose}).  Thus,
the last term in Eq.\ (\ref{iLorenzBose}) can be called the ``Bose
condensate''.  Our definition of the ``Bose condensate'' differs from
Ref.\ \cite{Bouchaud}, where it was associated with the case, where
the integral $y(r)$ diverges at the upper limit, and almost all income
is concentrated at the upper end.  We have never encountered such a
situation in the data.  We find that $y(r)$ always converges, and the
``Bose condensate'' fraction $f$ has a modest value.  The right panel
of Fig.\ \ref{fig:irs97G+P} demonstrates that formula
(\ref{iLorenzBose}) agrees very well with the data, giving $f=16\%$
for 1997.  However, Fig.\ 3 in Ref.\ \cite{DY-income} shows that $f$
monotonously increases in time for the last 20 years.  Wealth
distribution is also described by formula (\ref{iLorenzBose}) with
$f=16\%$ \cite{DY-wealth}.

\begin{figure}
\centerline{
  \epsfig{file=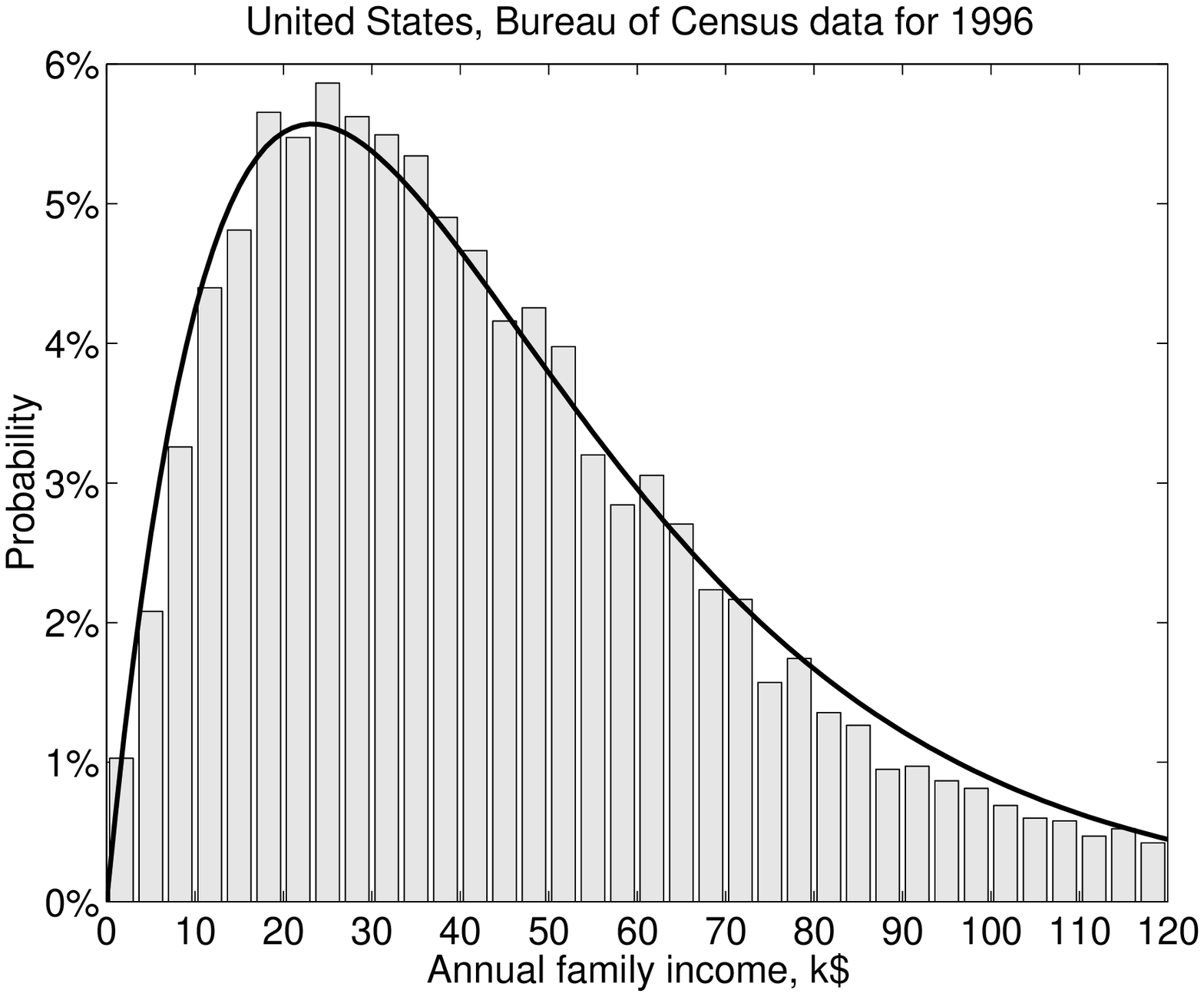,width=0.55\linewidth}
  \epsfig{file=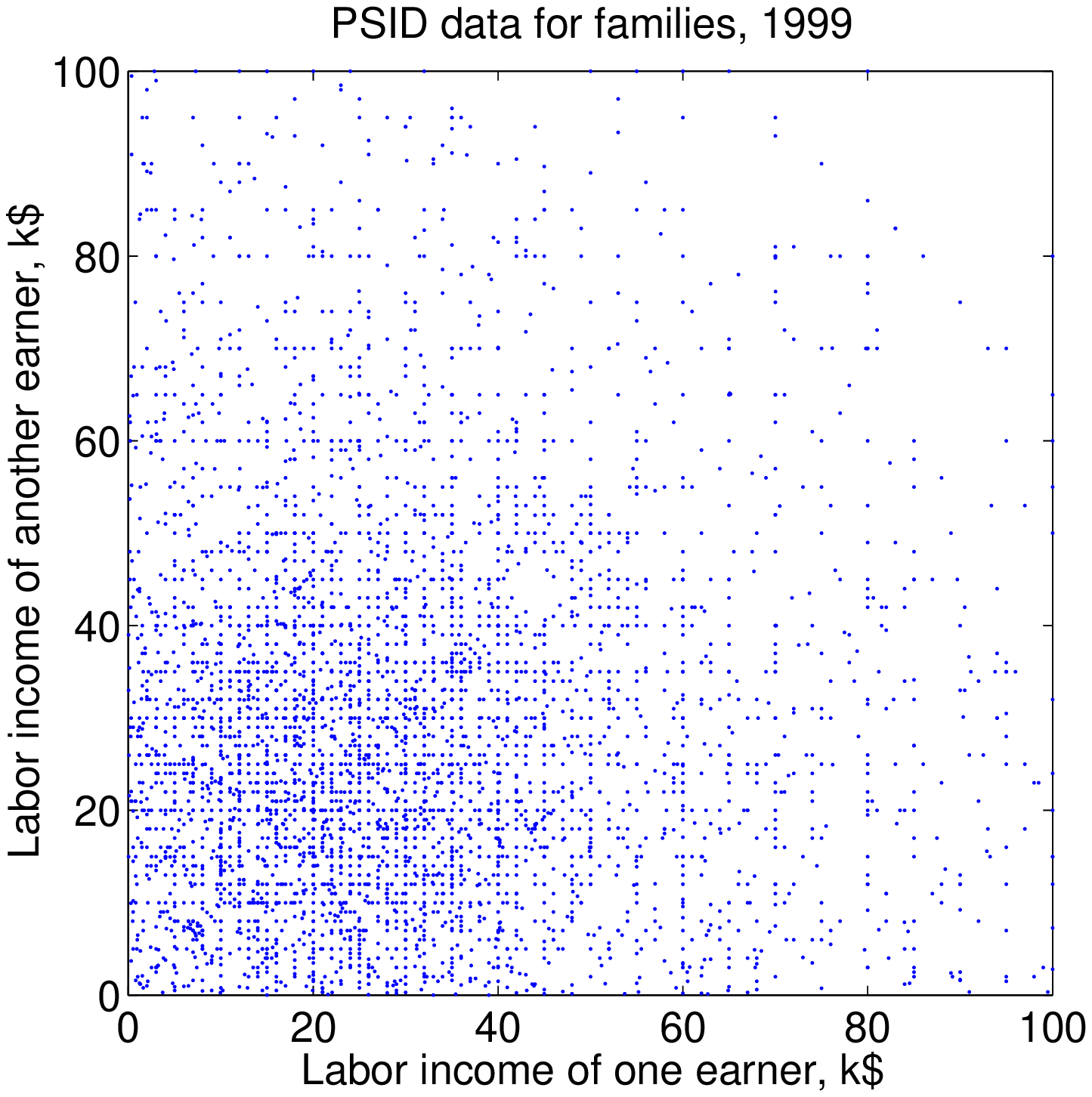,width=0.45\linewidth}}
\caption{Left panel: The histogram of the family income PDF for the
  families with two adults in 1996 \cite{DY-income,census} and its fit
  to Eq.\ (\ref{eq:family}).  Right panel: The incomes of spouses
  $(r_1,r_2)$ and $(r_2,r_1)$ for the families with two earners in
  1999 \cite{Michigan}.}
\label{fig:twoEarnersPSID}
\end{figure}

Now let us discuss the distribution of income for families with two
earners.  The family income $r$ is the sum of two individual incomes:
$r=r_1+r_2$.  Assuming that the individual incomes $r_1$ and $r_2$ are
uncorrelated and have exponential distributions, the family income PDF
$P_2(r)$ is given by the convolution
\begin{equation} \label{eq:family}
  P_2(r)=\int_0^{r}P_1(r')P_1(r-r')\,dr'= \frac{r}{R^2}e^{-r/R}.
\end{equation}
As shown in the left panel of Fig.\ \ref{fig:twoEarnersPSID}, Eq.\
(\ref{eq:family}) is in excellent agreement with the data.  The points
in the right panel show $(r_1,r_2)$ and $(r_2,r_1)$ for the families
in the data set.  It demonstrates that there is no significant
correlation between incomes of spouses.

\begin{figure}
\centerline{
  \epsfig{file=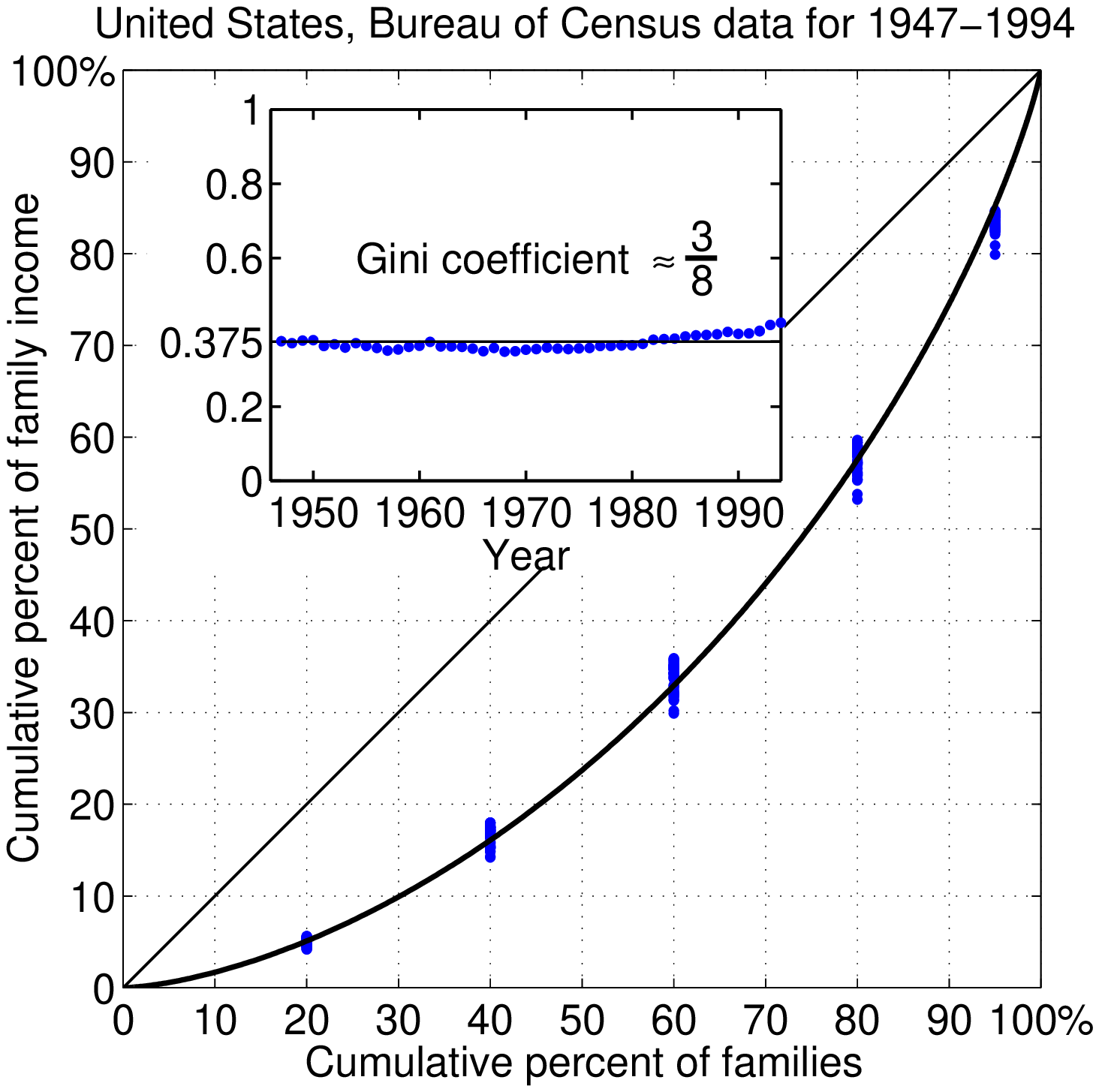,width=0.46\linewidth}
  \epsfig{file=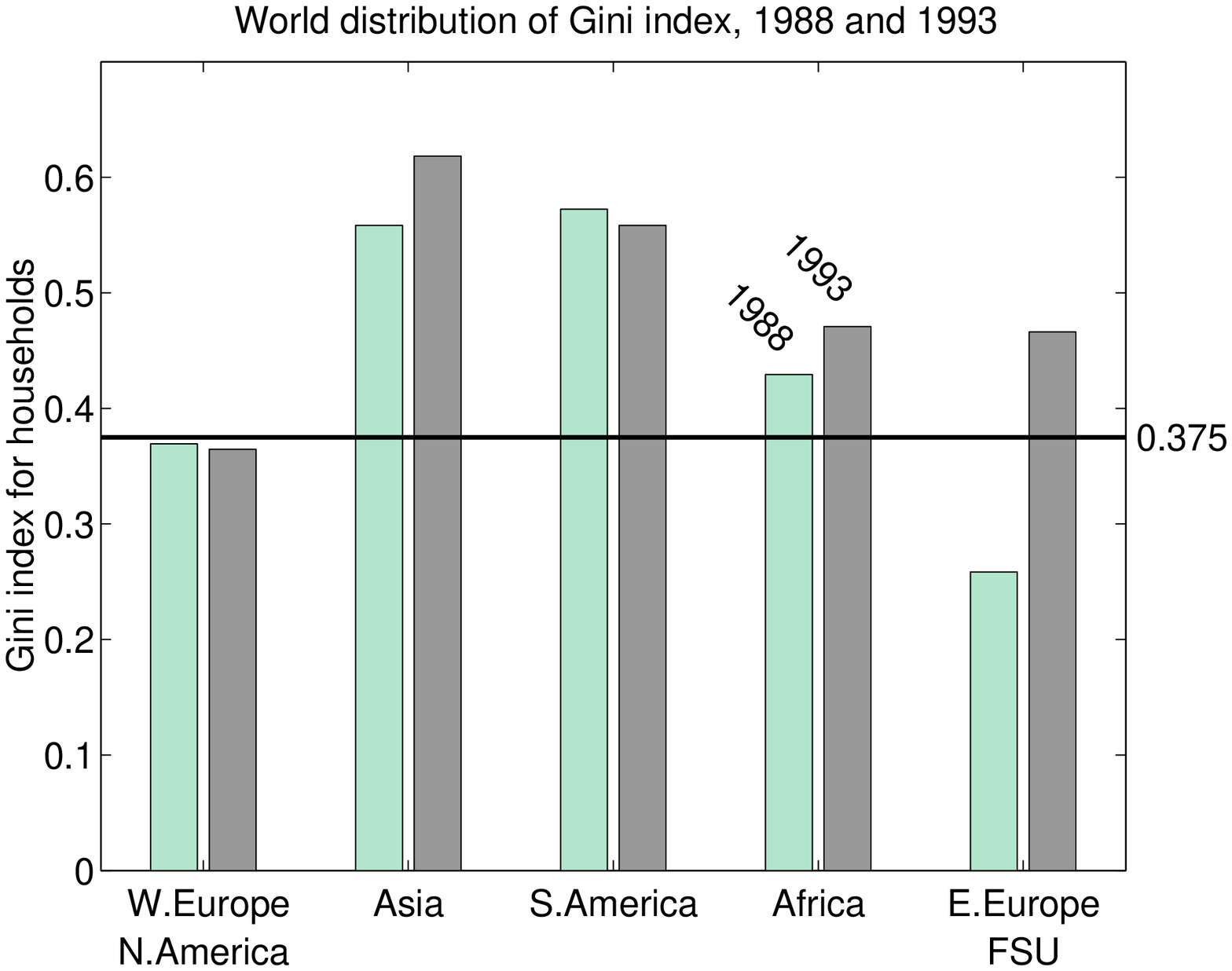,width=0.54\linewidth}}
\caption{Left panel: Lorenz plot calculated for the family income PDF
  (\protect\ref{eq:family}), compared with the US Census data points
  for families during 1947--1994 \cite{history}.  Inset: The US Census
  data points \cite{history} for the Gini coefficient for families,
  compared with the theoretically calculated value 3/8=37.5\%.  Right
  panel: Gini coefficients for households across the globe for two
  different years, 1988 and 1993 \cite{Milanovic}.}
\label{fig:giniWB}
\end{figure}

For the PDF $P_2(r)$ (\ref{eq:family}), the Lorenz curve was
calculated in Ref.\ \cite{DY-income}, and the Gini coefficient was
found to be $G_2=3/8=37.5\%$.  The left panel in Fig.\
\ref{fig:giniWB} demonstrates that these theoretical results are in
excellent agreement with the data for the last 50 years.  The right
panel shows the World Bank data \cite{Milanovic} for the average
values of the Gini coefficient in different regions of the world.  For
the well developed market economies of West Europe and North America,
the Gini coefficient is very close to the calculated value 37.5\% and
does not change in time.  In other regions of the world, the income
inequality is higher.  The special case is the former Soviet Union and
East Europe, where inequality was lower before the fall of communism
and greatly increased afterwards.  In statistical physics, the
exponential Boltzmann-Gibbs distribution is the equilibrium one,
because it maximizes the entropy of the system.  By analogy, we argue
that the equilibrium distribution of individual income in society is
also described by the Boltzmann-Gibbs law, and the equilibrium
inequality is characterized by the Gini coefficients $G_1=1/2$ for
individual income and $G_2=3/8$ for family income.  Fig.\
\ref{fig:giniWB} shows that such an equilibrium state of maximal
entropy has been achieved in developed capitalist countries.


\begin{thebibliography}{00}

\bibitem{DY-money} Dr\u{a}gulescu, A. A., and Yakovenko, V. M., 
  \textit{Eur. Phys. J. B} {\bf 17}, 723 (2000).

\bibitem{DY-income} Dr\u{a}gulescu, A. A., and Yakovenko, V. M., 
  \textit{Eur. Phys. J. B} {\bf 20}, 585 (2001).

\bibitem{DY-wealth} Dr\u{a}gulescu, A. A., and Yakovenko, V. M., 
  \textit{Physica A} {\bf 299}, 213 (2001).

\bibitem{Redner} Ispolatov, S., Krapivsky, P. L., and Redner, S., 
  \textit{Eur.  Phys. J. B} {\bf 2}, 267 (1998); Chakraborti, A., and
  Chakrabarty, B. K., \textit{Eur. Phys. J. B} {\bf 17}, 167 (2000);
  Braun, D., \textit{Physica A} {\bf 290}, 491 (2001).

\bibitem{Hayes} Hayes, B., \textit{American Scientist} {\bf 90}, 400 
  (2002).

\bibitem{Pub1304} \textit{Statistics of Income--1997, Individual Income
    Tax Returns}, Pub. 1304, Rev. 12-99, IRS, Washington DC, 1999;
  \url{http://www.irs.ustreas.gov/taxstats/}.

\bibitem{Bouchaud} Bouchaud, J.-P., and Mezard, M., \textit{Physica A} 
  {\bf 282}, 536 (2000).

\bibitem{census} The US Census data, \url{http://ferret.bls.census.gov/}.

\bibitem{Michigan} The PSID Web site, \url{http://www.isr.umich.edu/src/psid/}.

\bibitem{history} Weinberg, D.H., \textit{A Brief Look at Postwar U.S.
    Income Inequality}, P60-191, US Census Bureau, Washington, 1996;
    \url{http://www.census.gov/hhes/www/p60191.html}.

\bibitem{Milanovic} Milanovi\'c, B., World Bank Working Paper 2244, (1999);
  \url{http://www.worldbank.org/poverty/inequal/abstracts/recent.htm}.

\end{thebibliography}
\end{document}